\documentclass[journal]{IEEEtran}
\IEEEoverridecommandlockouts
\usepackage{cite}
\usepackage{graphicx}
\usepackage{epstopdf}
\usepackage{amsmath,amssymb,amsfonts}
\usepackage{algorithmic}
\usepackage{textcomp}
\usepackage{xcolor}
\usepackage{tabularx,booktabs}
\usepackage[flushleft]{threeparttable}

\usepackage{booktabs,ragged2e}
\usepackage{multirow}
\usepackage{subfigure}
\usepackage{bm}
\usepackage[colorlinks,
            linkcolor=red,
            anchorcolor=blue,
            citecolor=green
            ]{hyperref}
\def\BibTeX{{\rm B\kern-.05em{\sc i\kern-.025em b}\kern-.08em
    T\kern-.1667em\lower.7ex\hbox{E}\kern-.125emX}}
    
\begin{document}
    \title{Dilated Convolution based CSI Feedback Compression for Massive MIMO Systems}
    \author{Shunpu Tang, Junjuan Xia,  Lisheng Fan, Xianfu Lei, Wei Xu, and Arumugam Nallanathan, \IEEEmembership{Fellow, IEEE}
    \thanks{S. Tang,  J. Xia and L. Fan are all with the School of Computer Science, Guangzhou University, Guangzhou 510006, China (e-mail: tangshunpu@e.gzhu.edu.cn, \{xiajunjuan, lsfan\}@gzhu.edu.cn).}
    \thanks{X. Lei is with the Institute of Mobile Communications, Southwest Jiaotong University, Chengdu, China (e-mail: xflei@home.swjtu.edu.cn).}
    \thanks{W. Xu is with the National Mobile Communications Research Laboratory, Southeast University, Nanjing 210096, China. (e-mail: wxu@seu.edu.cn)}
    \thanks{A. Nallanathan is with the School of Electronic Engineering and Computer Science, Queen Mary University of London, London, U.K (e-mail: a.nallanathan@qmul.ac.uk).}
    \vspace*{-3mm}
    }

\maketitle
\begin{abstract}
Although the frequency-division duplex (FDD) massive multiple-input multiple-output (MIMO) system can offer high spectral and energy efficiency, it requires to feedback the downlink channel state information (CSI) from users to the base station (BS), in order to fulfill the precoding design at the BS. However, the large dimension of CSI matrices in the massive MIMO system makes the CSI feedback very challenging, and it is urgent to compress the feedback CSI. To this end, this paper proposes a novel dilated convolution based CSI feedback network, namely DCRNet. Specifically, the dilated convolutions are used to enhance the receptive field (RF) of the proposed DCRNet without increasing the convolution size. Moreover, advanced encoder and decoder blocks are designed to improve the reconstruction performance and reduce computational complexity as well. Numerical results are presented to show the superiority of the proposed DCRNet over the conventional networks. In particular, the proposed DCRNet can achieve almost the state-of-the-arts (SOTA) performance with much lower floating point operations (FLOPs). The open source code and checkpoint of this work are available at \href{https://github.com/recusant7/DCRNet}{https://github.com/recusant7/DCRNet}.

\end{abstract}

\begin{IEEEkeywords}
Massive MIMO, CSI feedback, deep learning, dilated  convolutions.  
\vspace*{-3mm}
\end{IEEEkeywords}
\section{Introduction}

As one of the key technologies for the next-generation wireless communications, massive multiple-input multiple-output (MIMO) systems can offer high spectral and energy efficiency by employing numerous antennas at both the transmitter and receiver. In MIMO system, the base station (BS) requires the downlink channel state information (CSI) to fulfill the precoding design. However, in the frequency-division duplex (FDD) mode, the downlink CSI has to be estimated by the user equipments (UEs), since it is difficult for the BS to obtain the CSI information due to the weak reciprocity. Notably, in massive MIMO systems, the overhead to feed back the huge CSI matrices increases significantly with the large number of antennas. Moreover, the transmit power and uplink bandwidth of users are limited as well. Hence, it is of vital importance to effectively compress the feedback CSI for the massive MIMO systems.

To solve this problem, traditional compressed sensing (CS) methods were proposed to compress the CSI matrix at the UEs and restore the information at the BS \cite{CS_feedback}. However, the CS based methods have several limitations, such as the ideal assumption of CSI sparsity, the neglect of channel statistics during the random project, and the iterative processing leading to a large latency. Recently, deep learning (DL) has attracted many research interests and shown the great successes in many fields. In this direction, a DL based CSI compression algorithm named CsiNet was proposed in \cite{wen_deep_2018}, which could outperform the conventional CS-based schemes. However, its two $3\times$3 convolutions failed to offer enough receptive field (RF), indicating that the neuron in deep layers could not represent enough region from the original input. Accordingly, a large size convolution such as $7\times 7$ even $9 \times 9$ was used to obtain a large RF in order to enhance the CSI reconstruction performance, at the cost of the increased computational complexity \cite{crnet,guo_convolutional_2020,lu_aggregated_2021,sun_ancinet_2020}.

In this paper, we investigate the problem of CSI feedback compression for the classical massive MIMO systems under FDD mode,
and propose a novel dilated convolution based CSI feedback network, namely DCRNet, which can help compress the CSI feedback efficiently and meanwhile reduce the computational complexity. In this deep network, the dilated convolutions are used to enhance the RF of the network without increasing the convolution size. Moreover, advanced encoder and decoders blocks are designed to improve the reconstruction performance and reduce computational complexity as well. Numerical results are finally presented to show the superiority of the proposed DCRNet over the conventional networks. In particular, the proposed DCRNet can achieve almost the state-of-the-arts (SOTA) performance with a much lower floating point operations (FLOPs).

\section{CSI Feedback in Massive MIMO Systems}
This paper investigates a single-cell FDD massive MIMO-OFDM system, where there are
$N_t \gg 1$ transmit antennas at the BS, and a single antenna at the user with $N_c$ sub-carriers. The complex-valued received signal at the $n$-th sub-carrier is
\begin{align}\label{receive signal}
    y_n=\bm{h}_{n}^H\bm{b}_n x_n+z_n,
\end{align}
where $\bm{h}_n \in \mathbb{C}^{N_t \times 1}$ is the channel gain vector, in which $(\cdot)^H$ denotes the conjugate transpose operation. Notation $\bm{b}_n \in \mathbb{C}^{N_t \times 1}$ represents the beamforming vector, $x_n$ is the transmitted symbol and $z_n$ denotes the additive white Gaussian noise (AWGN).  The CSI of all sub-carriers can be expressed by the matrix $\bm{H}=[\bm{h}_1,\bm{h}_2,\cdots,\bm{h}_{N_c}]$, where there are $2N_c N_t$ elements in total.

\begin{figure}[t!]
    \centering
    \includegraphics[width=3.5in]{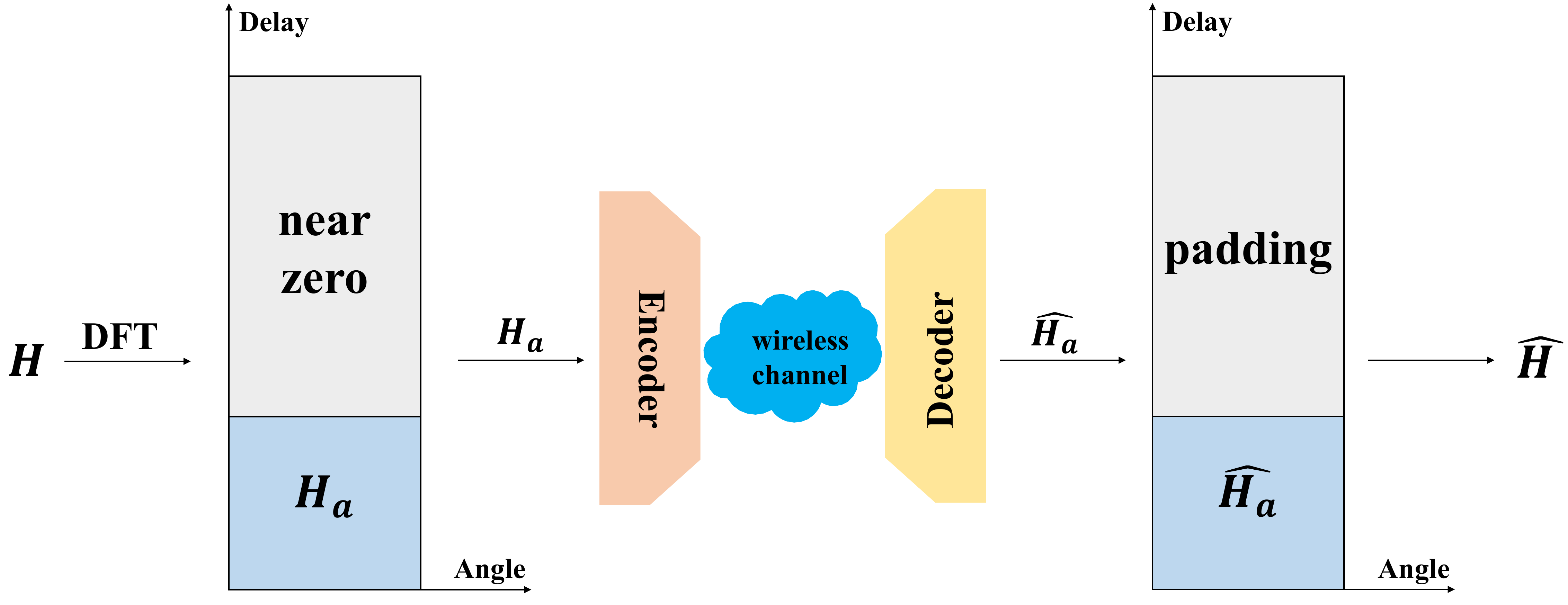}
    \caption{Structure of the CSI feedback in massive MIMO system.}
    \label{Fig::overview}
\end{figure}

The beamforming and precoding design at the BS requires the feedback of the CSI matrix $\bm{H}$, which however causes a severe communication overhead due to a large number of elements, especially in the practical massive MIMO systems. To solve this problem, the CSI matrix $\bm{H}$ should be compressed and fed back to the BS, where the CSI feedback model is shown in Fig. \ref{Fig::overview}. Specifically, the CSI matrix $\bm{H}$ is firstly transformed into the angular and delay domains, by using the discrete Fourier transformation (DFT), given by
\begin{align}\label{DFT}
    \bm{\widetilde{H}}=\bm{F}_c \bm{H} \bm{F}_t ,
\end{align}
where $\bm{F}_c$, $\bm{F}_t$ are $N_c \times N_c$ and $N_t \times N_t$ DFT transformation matrices, respectively.  The matrix $\bm{\widetilde{H}}$ is sparse and compressible, and it can be divided into two parts. One part is $\bm{H}_a$ which has $N_a$ rows composed of non-zero elements ($N_a<N_c$), while the other part contains the rest $N_c-N_a$ rows composed of near-zero elements. In this way, we can compress the CSI matrix $\bm{H}$, by using the matrix $\bm{H}_a$, with a negligible information loss.

To further compress the CSI matrix, deep autoencoder based compression and recovery strategies should be used, where an encoder is
employed in Fig. 1 to compress the CSI and output a codeword vector with a smaller size, given by
\begin{align}
    \label{encoder}
    \bm{v}=\mathcal{E}(\theta_1,\bm{H_a}),
\end{align}
in which $\mathcal{E}(\cdot)$ represents the compression operation and $\theta_1$ denotes the encoder's parameters. The compressed CSI will be fedback to the BS via the dedicated channel, and the BS can recover the received CSI matrix through a decoder, given by
\begin{align}
    \label{decoder}
    \bm{\hat{H_a}}=\mathcal{D}(\theta_2,\bm{v}),
\end{align}
where $\mathcal{D}(\cdot)$ and $\theta_2$ represent the recovery operation and model parameters, respectively. The aim of the CSI feeback is to minimize the average reconstruction error, given by
\begin{align}
    \label{object}
    \min_{(\theta_1,\theta_2)} \mathbb{E}||\bm{H_a}-\bm{\hat{H_a}}||_2.
\end{align}

\begin{figure*}[t]
    \centering
    \includegraphics[scale=0.9]{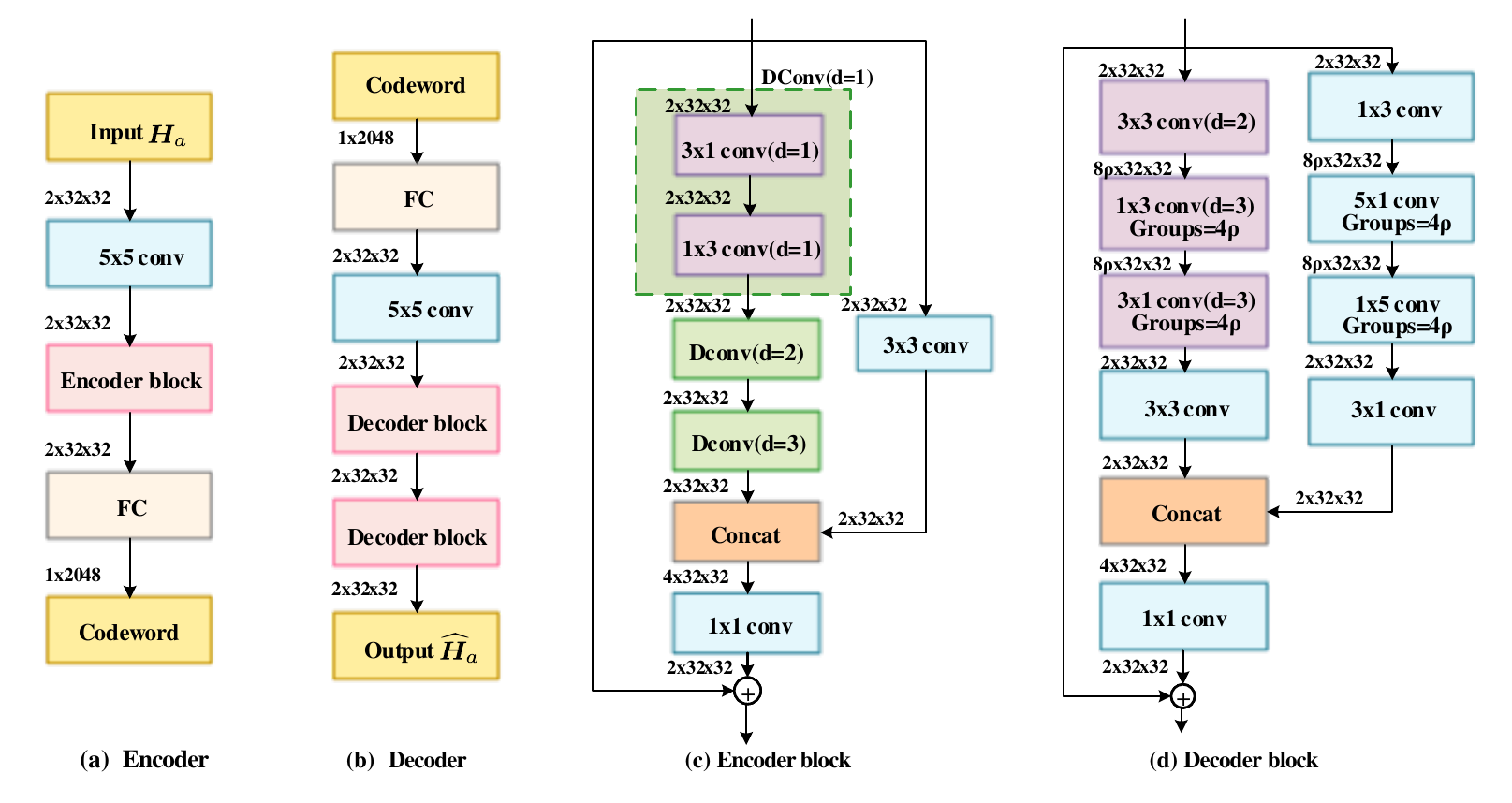}
    \caption{Architecture of the proposed DCRNet, where (a) and (b) illustrate the structures of the encoder and decoder of DCRNet. In addition,  the structures of the encoder block and decoder block in (a) and (b) are illustrated in (c) and (d) with $N_a=N_t=32$, respectively. In general, the encoder block and the decoder block are the main components of DCRNet, where BN layers, activation functions and reshape operations are omitted for brevity.}
    \label{Fig::DCRNet}
    \vspace*{-4mm}
\end{figure*}

\vspace{-3mm}
\section{Dilated convolutions based CSI feedback}
In this section, we will firstly describe the structure of the proposed DCRNet, and then give the critical design on the dilated encoder and decooder blocks which can extract features from the block-sparse CSI matrices efficiently.

\subsection{Structure of the proposed DCRNet}
Inspired by the fact that the CNN-based autoencoder can efficiently extract the spatially local correlation in the CSI matrices \cite{wen_deep_2018}, we devise the DCRNet by using a dilated CNN-based autoencoder, whose architecture is shown in Fig. 2. Specifically, the input matrix of the encoder is $\bm{H_a}$ with the dimension of $2 \times N_a \times N_t$, where the first two independent channels represent the real and imaginary parts of the CSI matrix, respectively. We then use a $5 \times 5$ head convolution to extract the features from the input CSI matrix and fuse the information from the real and imaginary parts. The output of the $5 \times 5$ head convolution is passed through an encoder block, which can help extract a deep abstract information. Unlike the ACRNet which uses two encoder blocks, the DCRNet only uses one encoder block to reduce the implementation complexity. After passing though the encoder block, the $2 \times N_a \times N_t$ will be reshaped to a 1-D vector, and the fully connected (FC) layers are used to compress the vector into a codeword with a compression rate of $\eta \in (0,1)$. After that, the codeword  will be transmitted to the decoder at the BS via the wireless channel.

When the decoder receives the codeword, it will restore the CSI matrix's dimension through the FC layers and a reshape operation. Then, a head convolution with a kernel size of $5 \times 5$ is employed to enhance the recovery performance. Afterwards, two sequential dilated decoder blocks are in charge of recovering the compressed information. Note that batch normalization (BN) and parametric rectified linear unit (PReLU) activation functions are appended to all the convolutions. The PReLU activation with a learnable parameter $\alpha$ can be expressed as
 \begin{align}
    \label{PReLU}
  \centering
\text{PReLU}(x)=\left\{\begin{array}{ll}
x, & x \geq 0 \\
\alpha x, & x<0.
\end{array}\right.
\end{align}

\subsection{Critical design on dilated encoder and decoder blocks}
To increase the RF for enhancing the CSI reconstruction performance and meanwhile avoiding the increase in the computational cost, we design a novel encoder and decoder blocks composed of a series of DConvs. Different from the standard convolution, Dconv extracts features with a specific interval $d$, which is also called as dilated rate. Formally, the 2D-dilated convolutional operation without bias can be written as
\begin{align}
    \label{dconv}
(\bm{I}\circledast \bm{K})[i,j]=\sum_u\sum_v {\bm{I}[i+du,j+dv]\cdot \bm{K}[u,v]},
\end{align}
where $\circledast$, $\bm{I}$ and $\bm{K}$ represent the DConv operator, 2D input and convolution kernel, respectively. In addition, $u$ and $v$ are the indices of convolution kernel $\bm{K}$.   The effective kernel size of DConv with dilated rate $d$ can be expressed as
\begin{align}
    \label{dconv_eqv}
k_i'=k_i+(k_i-1)(d-1),
\end{align}
where $k_i$ and $k_i'$ are the used and effective kernel sizes of convolution, receptively. Fig. 3 shows the dilated convolution with several dilated rates $d$. Specifically, when the dilated rate $d$ is 1, the dilated convolution degenerates into the standard convolution. In contrast, when $d>1$, the DConv operation can offer larger RF compared to the standard convolution with the same kernel size. This can achieve a sparse sampling for the block-sparse CSI matrix. In the following, we will describe the design details of the dilated encoder and decoder.

\subsubsection{Encoder block design}
Although the RF in the Dconv becomes larger with an increased dilated rate, the sequential use of Dconvs with equal dilated rates will cause the gridding effect and information loss \cite{WACV}.  Inspired by the technique of hybrid dilated convolutions (HDC) \cite{WACV} and asymmetric convolution \cite{crnet},  this paper employs asymmetric $3\times$3 convolutions with $d=1$, $d=2$ and $d=3$ in the encoder block, as shown in Fig. \ref{Fig::DCRNet}. One can readily verify that the proposed encoder block can obtain RF with the size of $13 \times 13$, which is equal to the size in CsiNet$+$ \cite{guo_convolutional_2020}. This can not only help the encoder block extract large-scale information, focus on the local details without occurring the gridding effect, but also decrease the computational complexity.

Since concatenating the features from different convolutions can yield the multi-resolution of CSI to enhance the system performance, we next perform the concatenation operation for the proposed encoder block. Specifically, we adopt a standard $3 \times 3$ filter as a complement to the Dconv, where the input of the encoder block will pass through two parallel branches. In particular, one branch is the aforementioned Dconv, while the other branch is a standard $3 \times 3$ convolution. Then, the outputs are concatenated with the size of $4\times N_a \times N_t$ and the number of channels will be reduced to 2 by the last $1\times 1$ convolution. In further, the residual learning is used, so that the identity of the input can be added to the output of the last $1\times 1$ convolution.

\subsubsection{Decoder block design}
Following the design principle of the encoder block, we design the decoder block by using  two parallel branches and an identity map. Specifically, in the first branch, a $3\times 3$ dilated convolution with $d=2$ is used to increase the feature dimension to $8\rho$, where $\rho \geq 1$ is the network width expansion rate. By adjusting the expansion rate $\rho$, we can make DCRNet-$\rho \times$ suitable for devices with different computational capacities. Then, group convolutions are used where $3\times 1$ and $1\times 3$ DConvs with $d=3$ replace the standard $9\times 9$  convolution, which can reduce the computational complexity of the decoder significantly. Afterwards, a $3\times 3$ convolution is used to reduce the feature dimension to $2$.
In the second branch, we  use the group convolution and width expansion operation, instead of using only one standard convolution in the second branch of encoder blocks. This is because that the computational resources of BS are usually not as limited as UEs and width expansion operation is important for CSI feedback as well\cite{lu_aggregated_2021}. In detail, $1 \times 3$ and  $3 \times 1$ filters are used to increase and reduce the feature dimensions, respectively. $5 \times 1$ and $1\times 5$ group convolutions are similar to those in the first branch. The outputs of the two branches will be concatenated together and passed through a $1\times 1$ convolution. Due to the use of Dconv and multi-resolutions, DCRNet is much wider than ACRNet\cite{lu_aggregated_2021} with lower FLOPs. This makes the decoder of the proposed DCRNet achieve a fine trade-off between the performance and computational complexity.
\section{Numerical Results and Analysis}
\subsection{Experimental settings}
\subsubsection{Data generation}
Following the experimental setting in \cite{wen_deep_2018}, the system model of COST 2100 \cite{cost2100} is applied to obtain the training and test samples, where we use two types of scenarios including indoor with 5.3GHz and outdoor with 300MHz bandwidth.
There are $N_t=32$ uniform linear array (ULA) antennas as the BS and $N_c=1024$ sub-carries. Accordingly, the original $2\times32\times 1024$ CSI matrix is transformed into the angular-delay domain and then be truncated to the $2\times 32\times 32$ matrix.
The number of training, validation and test sample are 100,000, 30,000 and 20,000, respectively.
\begin{figure}[t!]
    \centering
    \includegraphics[scale=0.45]{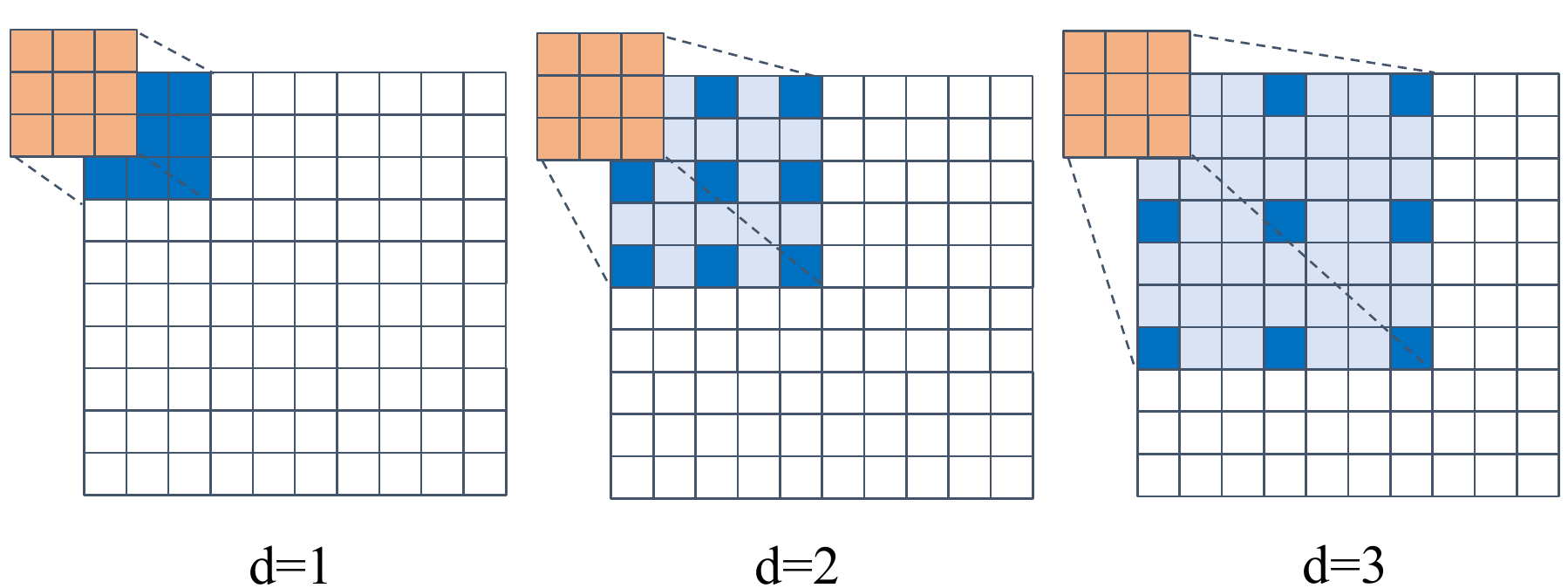}
    \caption{Demonstations of the dilated convolution operations with different dilated rates, where solid and shadow areas represent the effective operations and RF, respectively. When $d = 1$, the dilated convolution degenerates into the standard convolution. When $d> 1$, the dilated convolution is able to obtain a larger RF while it still involves the same computational complexity as the standard convolution.}
    \label{Fig::dilated_conv}
\end{figure}
\subsubsection{Training setting}
We use the Kaiming initialization to generate weights for each convolution layer and FC layer, where the Adam optimizer   is used to update the weights. Moreover, cosine annealing learning rate is exerted to improve
the performance according to \cite{crnet,lu_aggregated_2021}. In addition, the initial learning rate varies between $\gamma_{min}=5e-5$ and $\gamma_{max}=2e-3$,  and the process can be expressed as
\begin{align}
    \label{lr}
 \gamma_t =\gamma_{min}+\frac{1}{2}(\gamma_{max}-\gamma_{min})\Big(1+\cos\Big({\frac{t-T_w}{T-T_{w}}\pi}\Big)\Big),
\end{align}
where $t$ is the index of training epoch, $\gamma_t$ is the current learning rate, and $T_w$ and $T$ are the number of warm up epochs and total epochs, respectively. In this paper, $T_w$ and $T$ are set to 30 and 2500, respectively.
\subsubsection{Evaluation criteria}
We use normalized mean square error (NMSE) between the input CSI matrix $\bm{H_a}$ and the output matrix $\bm{\hat{\bm{H_a}}}$ restored by decoder to  evaluate the performance, which is given by
\begin{align}
    \label{nmse}
\mathrm{NMSE}=\mathbb{E}\left\{{\left\|\bm{H_a}-\hat{\bm{H_a}}\right\|_{2}^{2}}\big/{\left\|\bm{H_a}\right\|_{2}^{2}}\right\}.
\end{align}
In addition, the FLOPs and number of parameters are also measured in the experiment.
\subsection{Performance of the proposed DCRNet}
We compare the proposed DCRNet with the conventional DL-based CSI feedback networks, which are divided into the following two groups,
\begin{itemize}
    \item Low-complexity networks, such as the famous CsiNet\cite{wen_deep_2018}, CRNet\cite{crnet}, ACRNet-$1\times$ \cite{lu_aggregated_2021} and the proposed DCRNet-$1\times$. The FLOPs of this group are all less than 5M.
    \item High-performance networks, such as the proposed DCRNet-$10\times$, DS-NLCsiNet\cite{yu_ds-nlcsinet_2020}, the SOTA  CsiNet$+$\cite{guo_convolutional_2020} and ACRNet-$10\times$\cite{lu_aggregated_2021}. The FLOPS in this group are much higher than those in the first group.

\end{itemize}

\begin{table}[ht]
\begin{threeparttable}
\caption{Comparison of NMSE and complexity of different methods.}
\label{tab:1}
\setlength\tabcolsep{0pt} 

\begin{tabular*}{\columnwidth}{@{\extracolsep{\fill}} cl cccc}
\toprule
     & &
     \multicolumn{2}{c}{Complexity} &
     \multicolumn{2}{c}{NMSE (dB)} \\
\cmidrule{3-6}
     $\eta$ & Methods & FLOPs & Parameters   & Indoor & Outdoor\\
\midrule
    \multirow{8}{*}{1/4} &
     CsiNet &5.41M &2103K  &-17.36 &-8.75  \\
      & CRNet &5.12M &2103K  &-26.99 &\textbf{-12.70}  \\
        & ACRNet-$1\times$ &4.64M &2102K  &-27.16 &-10.71  \\
         &  DCRNet-$1\times$(ours)  &\textbf{4.01M} &\textbf{2102K}  & \textbf{-28.04}& -12.58 \\
         \cmidrule{2-6}
       & DS-NLCsiNet &\textbf{11.30M} &\textbf{2108K}  &-24.99 &-12.09  \\
        &CsiNet$+$ &24.57M &2122K  &-27.37 &-12.40  \\

      & ACRNet-$10\times$ &24.40M &2123K  &-29.83 &-13.61  \\

        &  DCRNet-$10\times$(ours)  &17.57M &2115K  & \textbf{-30.61}& \textbf{-13.72}\\
\midrule
    \multirow{8}{*}{1/8} &
        CsiNet &4.37M &1054K  &-12.70 &-7.65  \\
         & CRNet &4.07M &1054K  &-16.01 &\textbf{-8.04}  \\
          & ACRNet-$1\times$ &3.60M &1054K  &-15.34 &-7.85  \\
      & DCRNet-$1\times$(ours)  &\textbf{2.96M} &\textbf{1053K}  & \textbf{-16.26}& -7.95 \\
        \cmidrule{2-6}
       & DS-NLCsiNet &\textbf{10.25M} &\textbf{1059K}  &-17.00 &-7.96  \\
        &CsiNet$+$ &23.52M &1073K  &-18.29 &-8.72  \\
         & ACRNet-$10\times$ &23.36M &1074K  &-19.75 &-9.22  \\
         & DCRNet-$10\times$(ours) &16.52M &1066K  &-\textbf{19.92} &\textbf{-10.17}  \\

\midrule
    \multirow{8}{*}{1/16} &
     CsiNet &3.84M &530K  &-8.65 &-4.51  \\
      & CRNet &3.55M &530K  &-11.35 &-5.44  \\
      & ACRNet-$1\times$ &3.07M &529K  &-10.36 &-5.19  \\
       & DCRNet-$1\times $(ours)  &\textbf{2.44M} &\textbf{528K}  & \textbf{-11.74}& \textbf{-5.60} \\
        \cmidrule{2-6}
       & DS-NLCsiNet &\textbf{9.72M} &\textbf{534K}  &-12.93 &-4.98  \\
        &CsiNet$+$ &23.00M &549K  &-14.14 &-5.73  \\
         & ACRNet-$10\times$ &22.82M &549K  &\textbf{-14.32} &-6.30  \\
       & DCRNet-$10\times $(ours)  &16.00M &542K  & -14.02& \textbf{-6.35} \\

\midrule
    \multirow{8}{*}{1/32} &
     CsiNet &3.58M &268K  &-6.24 &-2.81  \\
      & CRNet &3.28M &267K  &-8.93 &\textbf{-3.51}  \\
        & ACRNet-$1\times$ &2.81M &267K  &-8.60 &-3.31  \\
       & DCRNet-$1\times$(ours)  &\textbf{2.18M} &\textbf{266K}  & \textbf{-9.05}& -3.47 \\
           \cmidrule{2-6}
            & DS-NLCsiNet &\textbf{9.46M} &\textbf{272K}  &-8.64 &-3.35  \\
        &CsiNet$+$ &22.74M &286K  &-10.43 &-3.4  \\
            & ACRNet-$10\times$ &22.50M &287K  &\textbf{-10.52} &-3.83  \\
              & DCRNet-$10\times$(ours) &15.74M &279K  &-9.88 &\textbf{-3.95}  \\

\bottomrule
\end{tabular*}

\smallskip
\scriptsize
\end{threeparttable}
\end{table}
Table \ref{tab:1} lists the performance comparison of the mentioned several CSI feedback networks, where the compression rate $\eta$ is set to $1/4$, $1/8$, $1/16$ and $1/32$, respectively. All the best results are presented in \emph{\textbf{bold}} font. From the first group of Table \ref{tab:1}, we can find that the proposed DCRNet-$1\times$ has the lowest computation complexity and fewest parameters. In particular, when the compression rate is 1/4, the proposed DCRNet-$1\times$ can reduce the FLOPs of CsiNet, CRNet and ACRNet-$1\times$ by 26\%, 21\%, 12\%, respectively. Moreover, the proposed DCRNet-$1\times$ can achieve the SOTA performance for various compression rates under indoor scenario.  Under outdoor scenario, the proposed DCRNet-$1\times$
can still achieve almost the SOTA performance with the lowest FLOPs.

From  the second group of Table \ref{tab:1}, we can find that the proposed DCRNet-$10\times$ has the second lowest computational complexity, and its FLOPs are about $7$M less than the SOTA CsiNet$+$ and ACRNet. Moreover, the proposed DCRNet-$10\times$ outperforms the other networks when the compression rates are $1/4$ and $1/8$ under both indoor and outdoor scenarios. With high compression rate such as $1/16$ and $1/32$, the proposed DCRNet-$10\times$ can outperform the other networks under outdoor scenario, and it can also achieve almost the SOTA result with a marginal performance gap of less than 0.7dB under indoor scenario. These results in Table I demonstrate that the proposed DCRNet-$1\times$ and DCRNet-$10\times$ can achieve almost the SOTA performance with  much lower FLOPs.
\begin{figure}[htp]
    \centering
    \subfigure[$\eta$=1/4]{
    \includegraphics[scale=0.48]{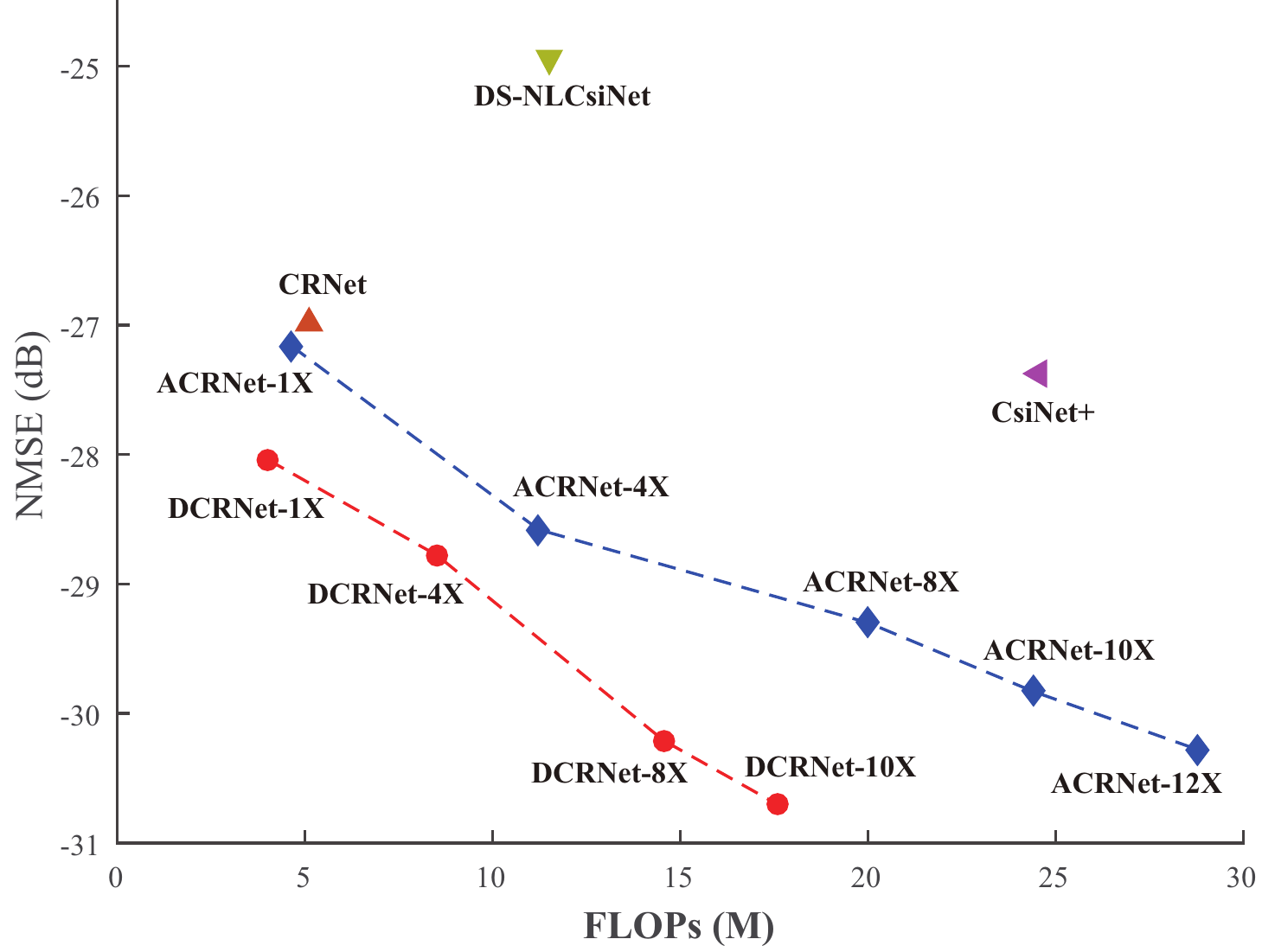}
    \label{Fig::flops_vs_nmse_4}
    }
    \subfigure[$\eta$=1/16]{
    \includegraphics[scale=0.49]{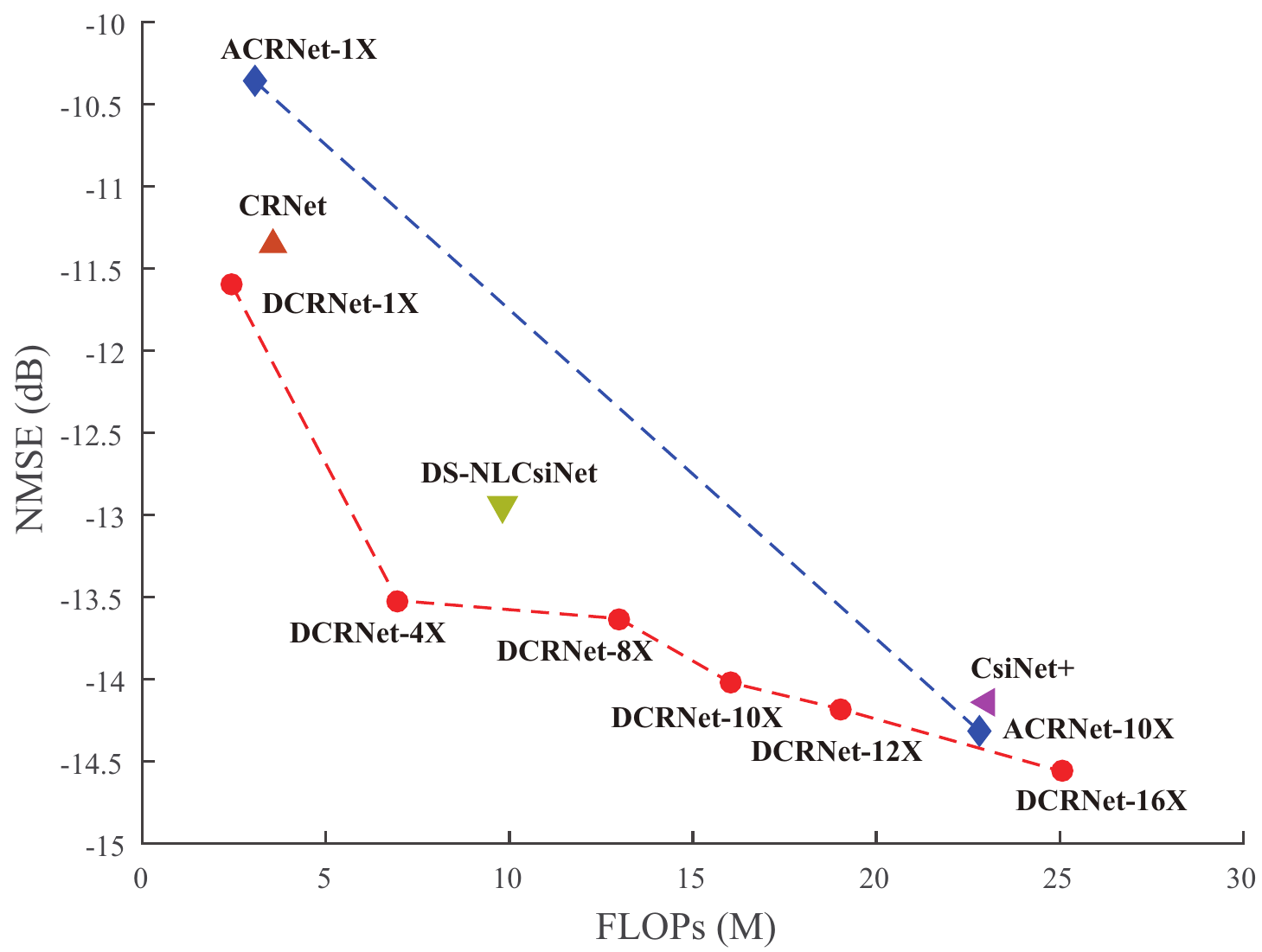}
    \label{Fig::flops_vs_nmse_16}
    }
    \caption{NSME performance versus FLOPs under indoor scenario when the compression rates are 1/4 and 1/16.}
    \vspace*{-5mm}
\end{figure}


Fig. \ref{Fig::flops_vs_nmse_4} depicts the channel reconstruction performance versus the  FLOPs, where the indoor scenario is investigated and $\eta$ is 1/4. In order to meet the need of the UEs with different computational capacities, the width expansion ratio $\rho$ of ACRNet and the proposed DCRNet are set to $1$, $4$, $8$, $10$ and $12$.  From Fig. \ref{Fig::flops_vs_nmse_4}, we can find that the curve of the proposed DCRNet is at the bottom left. Specifically, the DCRNet-$8\times$ achieves the same performance as the ACRNet- $12\times$ with only a half computational complexity. This demonstrates the high efficiency of the proposed DCRNet.

Fig. \ref{Fig::flops_vs_nmse_16} illustrates the NMSE performance versus FLOPs under indoor scenario, where the compression rate is 1/16 indicating that it is much more difficult to restore the CSI matrix from the codeword vector. From this figure, we can find  that the curve of the proposed DCRNet is still at the bottom left, which further demonstrates the advantages of the DCRNet.  In particular, compared with the CRNet and ACRNet-$1\times$, the DCRNet-$1\times$ can achieve the best performance with the lowest computational complexity. Moreover,  the proposed DCRNet-$4\times$ achieves -13.52dB NSME and 6.69M FLOPs,
which outperforms the -12.93dB NSME and 4.98M FLOPs of DS-NLCsiNet. In further, the proposed DCRNet-$8\times$ and DCRNet-$10\times$ achieve -14.02dB and -14.18dB NSME, respectively, and there is a marginal gap between them and -14.32dB NSME of the ACRNet-$10\times$. However, the FLOPs of DCRNet-$8\times$ and DCRNet-$10\times$ are much lower than those of the ACRNet-$10\times$. In a word, by adjusting the width expansion rates, the proposed DCRNet can achieve a fine trade-off between the NSME performance and computational complexity.

\subsection{Ablation experiments}
In order to further demonstrate the validity of the dilated convolution and large receptive filed, we replace the original dilated convolution in the proposed DCRNet. In particular, we use the standard convolution to replace all the dilated convolutions in both the encoder and decoder as a baseline. Besides, we  use the dilated convolution in the encoder only, to generate another network called by DCRNet-M1. The NMSE comparison results are given in Table \ref{tab:2}, where the compression rate is $1/16$. From this table, we can find that the proposed DCRNet can achieve the best reconstruction performance thanks to the joint use of dilated convolution in both the encoder and decoder. Moreover, the DCRNet-M1 outperforms the baseline, due to the use of dilated convolution in the encoder. 

\begin{table}[ht]

\begin{threeparttable}
\caption{Comparison of NMSE when compression rate is ${1}/{16}$.}
\label{tab:2}
\setlength\tabcolsep{0pt}
\begin{tabular*}{\columnwidth}{@{\extracolsep{\fill}} lccc}
\toprule
& Baseline & DCRNet-M1& DCRNet\\
\midrule
Indoor & -11.28 & -11.41& \textbf{-11.74}\\
Outdoor & -5.46 & -5.55& \textbf{-5.60}\\
\bottomrule
\end{tabular*}

\smallskip
\scriptsize
\end{threeparttable}
\end{table}

\section{Conclusions}
In this paper, we proposed a novel DL-based CSI feedback network for massive MIMO systmes named DCRNet, which 
employed dilated convolution to increase the RF and meanwhile reduce the computational complexity. 
Simulation results have been demonstrated to show the efficiency and flexiblity of DCRNet. Specifically, DCRNet-$1\times$ outperformed the competitive lightweight networks with limited computational resources. As well, DCRNet-$10\times$ achieved almost the SOTA reconstruction performance with a much lower computational complexity than the conventional networks.
\bibliographystyle{IEEEtran}

\bibliography{reference}
\end{document}